# Polarization-dependent ARPES study of $La_{1-x}Sr_xMnO_3$

M. C. Falub[1,♣], M. Shi[1,♦], P. R. Willmott[1], J. Krempasky[1], S. G. Chiuzbaian[1], K. Hricovini[2] and L. Patthey[1]

[1]Swiss Light Source, Paul Scherrer Institut, CH-5232 Villigen, Switzerland

[2]Laboratoire de Physique des Matériaux et des Surfaces, Université de Cergy-Pontoise, 95011 Neuville, France

*We present angle-resolved photoemission spectroscopy results on thin films of the three-dimensional manganite perovskite $La_{1-x}Sr_xMnO_3$, with $T_C = 240$ K. In the low-binding energy region, we found two energy bands, one dispersing along the (100) sample surface and the other along the normal to the surface. Through a detailed polarization-dependent analysis, we have determined the symmetries of the two bands. The initial states associated with the band that disperses along the surface are dominantly odd with respect to the (110) mirror plane and even with respect to the (010) mirror plane. Based upon this investigation, we have derived the expression of the spatial term for the wave functions describing these low-energy states.*

---

[♣] Present address: Ecole Polytechnique Fédérale de Lausanne (EPFL), Laboratoire de Spectroscopie Electronique, CH-1015 Lausanne, Switzerland.
[♦] Present address: Department of Physics, University of Illinois at Chicago, Chicago, Illinois 60607, USA.

*1. INTRODUCTION*

The discovery of the colossal magnetoresistance effect in manganite perovskites has stimulated considerable interest in their electronic and magnetic properties [1, 2]. The system $La_{1-x}Sr_xMnO_3$ (LSMO) has a very rich phase diagram [3, 4]. Within a certain range of doping, it shows a large decrease in resistivity upon cooling, associated with a transition from a paramagnetic (PM) to a ferromagnetic metallic (FM) state. Below the transition temperature ($T_C$), the resistivity is further strongly reduced by applying a magnetic field, known as colossal magnetoresistance (CMR). The phase transition from PM to FM and the resistivity dependence upon temperature have been qualitatively explained using the double-exchange (DE) model [5, 6], which assumes that in the FM phase, LSMO is a mixed-valence system with $Mn^{3+}$ ($t_{2g}^3 e_g^1$) and $Mn^{4+}$ ($t_{2g}^3$) ions. There is a density of (1-x) $e_g$ electrons per unit cell that are free to move through the crystal, subject to a strong Hund's coupling to the localized $Mn^{4+}$ (S = 3/2) spins. The kinetic (band) energy is minimized by making all spins parallel.

One major obstacle to understanding the physics of the manganites has been the lack of detailed information about their electronic structure, the low binding energy states and their associated wave functions, which are used to calculate important physical quantities. Angular-resolved photoelectron spectroscopy (ARPES) is a unique spectroscopic tool capable of revealing the **k**-resolved electronic structure [7, 8]. ARPES is a surface-sensitive technique. So far, the main difficulty encountered in using ARPES to study the three-dimensional (3D) manganites has been the preparation of smooth surfaces at an atomic level, because of the lack of a natural cleavage plane. At the Swiss Light Source, it has been possible to grow high-quality epitaxial thin films of LSMO with submonolayer roughness and investigate their **k**-dependent electronic structure by *in-situ* ARPES [9].

Here we report an *in-situ* ARPES study of thin films with $T_C$ = 240 K, close to the lowest transition temperature from paramagnetic to ferromagnetic metallic phase [3, 4]. Whereas a vanishing or very small spectral weight at the Fermi level ($E_F$) was reported for the similar layered (2D) manganites [10, 11], a finite step at $E_F$ was observed for the 3D-manganites.

Compared to our earlier ARPES results on similar films with a high transition temperature ($T_C$ = 313 K) [9], we show here that a similar sharp step at $E_F$ only appears in the 1D-like band. On the other hand, the 2D-like band lies always below $E_F$. We have varied the matrix elements by changing the polarization of the incoming light. In this manner, we have determined the symmetries of the two energy bands with respect to the reflections in different mirror planes and derived the spatial term of the corresponding wave functions.

2. EXPERIMENTAL

LSMO samples were prepared by growing *in-situ* 1300 Å-thick films heteroepitaxially on $SrTiO_3$ (001) substrates by a novel adaptation of pulsed laser deposition [12, 13]. *In-situ* reflection high-energy electron-diffraction patterns and Kiessig fringes in *ex-situ* X-ray reflectivity curves demonstrated that the final film had a surface roughness of less than one monolayer. Low-energy electron-diffraction analysis showed a clear (1x1) pattern with no sign of surface reconstruction. By controlling the Sr/La ratio in the deposition process, we could change the doping hole level and tune the transition temperature ($T_C$), accordingly. The sample used in this study has $T_C$ = 240 K, which is close to the lowest transition temperature from PM to FM states in this system.

ARPES measurements were performed at the Surface and Interface Spectroscopy (SIS) beamline at the Swiss Light Source [14, 15]. During measurements, the base pressure always remained below $1 \times 10^{-10}$ mbar. The ARPES spectra were recorded with a Gammadata Scienta 2002 analyzer. The energy resolution was relaxed to 40 meV to obtain a high photon flux. All measurements were performed at 20 K.

In our experimental set-up the propagation vector of the incident light, the surface normal to the sample and the direction of the outgoing photoelectrons lie in the same horizontal plane (HP). The vector potential (**A**) of the light with linear vertical polarization (LV) is perpendicular to HP, while it lies in HP and is 45° to the surface normal in the case of linear horizontal polarization (LH). The (010) and (110) mirror planes can be brought to HP by rotating the sample about the surface normal.

*3. RESULTS*

Figures 1a and 1b show the ARPES spectral intensity as a function of the wavevector (**k**) and binding energy ($E_b$) obtained in the (110) and (010) mirror planes, respectively. The corresponding paths in **k**-space are indicated in Fig. 1d. As the sample was rotated between these two planes (and using the appropriate polarization of the incoming light), the dispersion could be seen to monotonically evolve from Fig. 1a to Fig. 1b. This indicates that these dispersions belong to the same energy band, which has mainly 2D character, because the dispersion in a cut parallel to the (001) direction (not shown) is much weaker than those found for (110) and (010) [9].

We have also found the signature of a 1D-like band along the surface normal (001), similar to our previous study on samples with $T_C$ = 313 K [9]. Figure 1c shows that when the wave vector approaches the **k**-point (0, 0, 0.4 $\pi/a$) from both sides along the (001) direction, the step at $E_F$ becomes sharper. We observed a similar trend for any path parallel to (001) with $|k_{\parallel}| = (k_x^2+k_y^2)^{1/2} < 0.4\ \pi/a$. It should be noted that we found a non-vanishing intensity at $E_F$ only in the vicinity of $k_z$ = 0.4 $\pi/a$ and $|k_{\parallel}| < 0.4\ \pi/a$, where the 2D-like band disperses to higher binding energies. The finite spectral weight in the 1D-like band might explain why the 3D-manganites have a much higher $T_C$ than the layered manganites, for which only vanishing or very small spectral weights at $E_F$ were reported [10, 11].

The dispersive ARPES features strongly depend on the polarization of the photon beams. In the (110) mirror plane a clear peak disperses along a path parallel to the surface when using the LV light (Fig. 2a). However, when we used LH light, there is only a little residual intensity of the dispersive peak on a sloped background (Fig. 2b). Figure 2c shows the comparison of the energy distribution curves (EDCs) taken at the **k**-point (0.4, 0.4, 0.7) $\pi/a$ with LV and LH lights, respectively. An opposite behavior was found for the spectra taken in the (010) mirror plane (Fig. 2e and 2f). Here we observed a clear dispersive feature in the case of LH light, but none when using LV light. A comparison of the EDCs taken at **k** = (0.7, 0, 0.7) $\pi/a$ with LV and LH lights is presented in Fig. 2g.

The photoemission spectral weight is proportional to the square of the matrix element $M_{if}$ =<f|**A**_**p**|i>, where |i> and |f> are the initial and final states, and **p** is the momentum operator. In our experimental configuration, **A** is odd (even) with respect to HP for LV (LH) light. If HP is a mirror plane of the sample, then the final states in the

photoemission process must be even with respect to the reflection in the same mirror plane. Thus, to have non-zero matrix elements, the symmetry of **A** and that of the initial state |i> must be the same [16, 17]. Namely, |i> must be odd with respect to the mirror plane for LV light, and even for LH light. Although it is difficult to determine whether there are some residual dispersive features in Fig. 2b and 2e, our polarization-dependent analysis indicates that the symmetries of the initial states associated with the dispersion along the sample surface are dominantly odd with respect to the (110) mirror plane and even with respect to the (010) mirror plane.

## 4. WAVEFUNCTIONS

In a crystal with cubic symmetry the Mn 3d states are split into the doubly degenerated $e_g$ states and triply degenerated $t_{2g}$ states. For an octahedral environment of anions, like in LSMO, the $t_{2g}$ states have lower energies than the $e_g$ states [18]. When the symmetry is lower than cubic the $e_g$ states will split into different energy levels. Early studies of LSMO indicate that the $t_{2g}$ states are about 2eV below $E_F$, while the low binding-energy states near $E_F$ mainly originate from the $e_g$ states [19].

In order to analyze the wave functions of the low binding-energy states, let us consider $e_{g1}$ and $e_{g2}$ as basis functions of the $e_g$ states. The two basis functions are isomorphic to that of the cubic harmonics $3z^2 - r^2$ and $x^2 - y^2$, with the x, y and z axes along the [100], [010] and [001] directions, respectively. Other equivalent pairs of the basis functions can be generated through linear combinations. The normal to the sample surface is [001]. $e_{g1}$ possesses even symmetry with respect to both the (010) and (110) mirror planes, while $e_{g2}$ is even with respect to (010) and odd to (110). The Bloch sum of $e_{g1}$ can be constructed for a wave vector **k** as:

$$E_{g1} = N^{-1/2} \sum \exp(i\mathbf{k}\cdot\mathbf{R}) \, e_{g1}(\mathbf{r}-\mathbf{R}),$$

where N is the total number of lattice points and the summation runs over all the lattice points **R**. Similar linear combinations $E_{g2}$ and $O_{Pn}$ can be obtained from $e_{g2}$ and oxygen 2p orbitals. In the tight-binding scheme, the wave functions $\psi_1$ and $\psi_2$ for the low-binding energy states can be written as:

$$\psi_1 = A_1(\mathbf{k}) E_{g1} + B_1(\mathbf{k}) E_{g2} + \sum_n C_n(\mathbf{k}) O_{Pn} \quad (1)$$

$$\psi_2 = A_2(\mathbf{k}) E_{g1} + B_2(\mathbf{k}) E_{g2} + \sum_n D_n(\mathbf{k}) O_{Pn} \quad (2),$$

where A-D are **k**-dependent coefficients.

Let us consider $\psi_1$ and $\psi_2$, the wave functions associated with the dispersions observed along the sample normal and along its surface, respectively. When $\psi_2$ has exclusively odd parity with respect to the reflection in the (110) mirror plane, the terms $A_2(\mathbf{k}) E_{g1}$ and $\mathbf{P}_e \sum_n D_n(\mathbf{k}) O_{Pn}$ will be projected out from (2). Here $\mathbf{P}_e$ is the even symmetry projection operator with respect to the mirror plane. $\psi_2$ becomes:

$$\psi_2 = B_2(\mathbf{k}) E_{g2} + (1-\mathbf{P}_e) \sum_n D_n(\mathbf{k}) O_{Pn}. \quad (3)$$

As discussed above, the states associated with the dispersion observed in the mirror plane (110) have dominantly odd symmetry with respect to the reflection in this mirror plane. Therefore, their wave functions can be described as (2) with

$$|B_2(\mathbf{k})| \gg |A_2(\mathbf{k})| \quad (4)$$

$$|(1-\mathbf{P}_e) \sum_n D_n(\mathbf{k})| \gg |\mathbf{P}_e \sum_n C_n(\mathbf{k})|. \quad (5)$$

One can conclude that $\psi_2$ has mainly $x^2 - y^2$ character for those **k**-points that can be observed in the (110) mirror plane in ARPES measurements.

Since $\psi_1$ is orthogonal to $\psi_2$ for any **k**-point, if we neglect the overlap of the orbitals that belong to Mn atoms at different sites, $\psi_1$ becomes

$$\psi_1 \approx B_2(\mathbf{k}) E_{g1} - A_2^*(\mathbf{k}) E_{g2} + \sum_n C_n(\mathbf{k}) O_{Pn} + \ldots\ldots \quad (6)$$

According to (4), $\psi_1$ has mainly $3z^2 - r^2$ character and it is a linear combination of $e_{g1}$ and O 2p orbitals. The small residual intensity of the dispersive peak in Fig. 2b and 2e may results from small random deviations of the Mn orbitals from $x^2 - y^2$ and/or of the O 2p orbitals from the crystal axes.

Our ARPES results indicate that the symmetry of the studied system is lower than cubic. The $Mn^{3+/4+}$ ions are situated in an environment in which a large cubic field, which causes the splitting of the 3d states into $e_g$ and $t_{2g}$, is supplemented by a tetragonal field. The reasons are as following:

1) The electronic structure is anisotropic: we only found a non-vanishing intensity at $E_F$ in the vicinity of $(0,0,0.4)\pi/a$, but not near $(0.4,0,0)\pi/a$ and $(0,0.4,0)\pi/a$, which should be equivalent to $(0,0,0.4)\pi/a$ if the system would have cubic symmetry.

2) For cubic symmetry, the two bands resulting from the $e_g$ orbitals would be degenerate along the body diagonal Γ-R of BZ (the dashed line in Fig. 2d) and their wave functions would have different parities with respect to the reflection in the (110) mirror plane. However, we only found one energy band with odd symmetry to the (110) mirror plane in the vicinity of the intersection between Γ-R and the path along the spectra were taken (Fig. 2).

3) Because (110) and (010) remain mirror planes, the system has tetragonal symmetry with four-fold rotation axis along [001], which is the surface normal.

As a consequence of the lower symmetry, the doubly degeneracy of the $e_g$ states is lifted. The splitting of the $e_g$ levels is evident because no energy band crossover was observed at any **k**-point. It means that the energy band associated with the finite spectral weight at $E_F$ (Fig. 1c) lies above the other band which always remains below $E_F$ (e.g. Fig. 1a and 1b). Thus, the form for the wave functions we obtained from the ARPES data in the (110) mirror plane can be extended to general **k**-points in BZ.

This orbital assignment with the energy band agrees with the present experimental findings: the energy band associated with $\psi_2$ has mainly 2D-like character and disperses in the $(k_x, k_y)$ plane, which is parallel to the sample surface.

It is known that heteroepitaxial LSMO films grown on $SrTiO_3$ (001) substrates are under slight tensile stress due to lattice mismatch. Further studies are required to investigate the influence of this upon the electronic structure and the splitting of the $e_g$ levels.

In summary, the two-band picture [9] persists in the LSMO thin film with $T_C$ close to the lowest transition temperature from the PM to FM metallic state. We found a finite spectral weight at $E_F$ in the 1D-like band. Through a detailed polarization-dependent ARPES analysis, we have determined the symmetry of the 2D-like band. The initial states associated with this band are dominantly odd with respect to the (110) mirror plane and even with respect to the (010) mirror plane. Within the tight-binding scheme, we have derived the spatial-part of the wave functions that describe these two bands.

**Acknowledgements**

This work was performed at the Swiss Light Source, Paul Scherrer Institut, Villigen, Switzerland. We thank J. F. van der Veen and R. Abela for discussions and comments. R. Herger, M. Schneider and C.V. Falub are acknowledged for helping us to prepare and characterize our samples. This work was supported by the Paul Scherrer Institut.

**Figure captions**

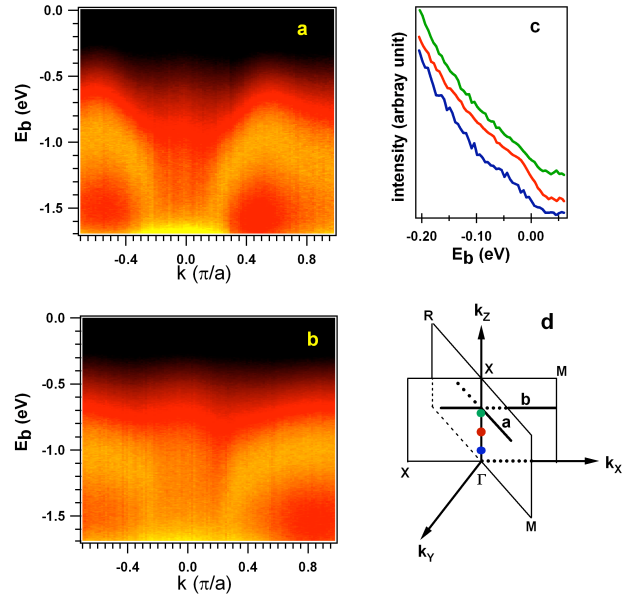

Figure 1: (a) and (b) ARPES intensity maps at 20 K; the corresponding paths in the Brillouin zone are indicated in Fig. 1d. (c) EDCs taken at several **k**-points (indicated by filled circles in Fig. 1d).

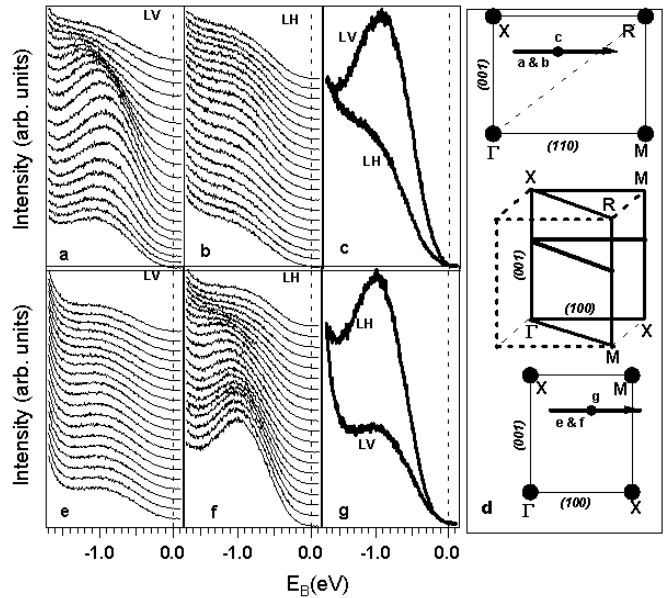

Figure 2: ARPES spectra taken at 20 K with linear vertical (LV) and linear horizontal (LH) light, respectively. The paths in the Brillouin zone corresponding to the spectra shown in Figs. 2 (a), (b), (e) and (f) are indicated by arrows in Fig. 2d**.** The **k**-points where the EDCs shown in Figs. 2 (c) and (g) were measured are marked in Fig. 2d by filled circles.